\documentclass[aps,prl,twocolumn,superscriptaddress]{revtex4-2}
\usepackage{graphicx,amssymb,amsfonts,amsmath,chemarr,color,commath,braket,hyperref,bm,bbold,dsfont}

\newcommand{\beginsupplement}{%
\setcounter{page}{1}
        \setcounter{equation}{0}
        \renewcommand{\theequation}{S\arabic{equation}}%
        \setcounter{table}{0}
        \renewcommand{\thetable}{S\arabic{table}}%
        \setcounter{figure}{0}
        \renewcommand{\thefigure}{S\arabic{figure}}%
     }

\begin{document}

\title{General mechanism for concentration-based cell size control}

\author{Motasem ElGamel}
\affiliation{Department of Physics and Astronomy, University of Pittsburgh, Pittsburgh, Pennsylvania 15260, USA}

\author{Lucas Ribaudo}
\affiliation{Department of Physics and Astronomy, University of Pittsburgh, Pittsburgh, Pennsylvania 15260, USA}

\author{Andrew Mugler}
\email{andrew.mugler@pitt.edu}
\affiliation{Department of Physics and Astronomy, University of Pittsburgh, Pittsburgh, Pennsylvania 15260, USA}

\begin{abstract}
Cells control their size to cope with noise during growth and division. Eukaryotic cells exhibiting “sizer” control (targeting a specific size before dividing) may rely on molecular concentration thresholds, but simple implementations of this strategy are not stable. We derive a general criterion for concentration-based sizer control and demonstrate it with a mechanistic model that resolves the instability by using multistage progression towards division. We show that if molecular dynamics in one stage satisfy the sizer criterion, then sizer control follows for the whole progression. We predict that perturbations to the molecular dynamics in non-sizer stages shift the size statistics without disrupting sizer control, consistent with recent experiments in fission yeast.
\end{abstract}

\maketitle

Size is an essential variable for cellular function across all organisms \cite{ginzberg2015being,cook2007size, amodeo2016cell,miettinen2016cellular}. It influences key processes such as nutrient intake \cite{turner2012cell, chien2012cell}, gene expression \cite{chen2020differential}, maintaining tissue uniformity \cite{ginzberg2018cell,ginzberg2015being}, metabolism \cite{miettinen2016cellular,cadart2022scaling}, and more \cite{marshall2012determines,young2006selective}. As cells are affected by intrinsic and extrinsic noise sources that influence their growth and division, they experience size fluctuations \cite{wang2010robust, susman2018individuality, tanouchi2015noisy}. Thus, they must maintain control over their size by adjusting the cell cycle in a size-dependent manner \cite{susman2018individuality,tanouchi2015noisy}. Although passive size control is sufficient in linearly growing cells, active size control is essential for stability in exponentially growing cells \cite{turner2012cell}. Different size control strategies have been identified, namely the sizer, adder, and near-timer \cite{amir2014cell,facchetti2017controlling,willis2017sizing,sauls2016adder}. While it is widely reported that bacteria employ the adder strategy, i.e., adding a constant size before dividing \cite{taheri2015cell, si2019mechanistic, sauls2016adder,amir2014cell}, fission yeast \cite{facchetti2017controlling,sauls2016adder,turner2012cell}, smaller daughters of budding yeast \cite{talia2007effects}, and some mammalian cells \cite{varsano2017probing,xie2020g1} implement the sizer strategy, targeting a specific size before dividing. It remains an open question how the sizer strategy is achieved.

Multiple molecular mechanisms have been proposed for size control in budding and fission yeast \cite{turner2012cell,wood2015sizing}. A strong candidate mechanism relies on a molecule, or group of molecules, that accumulates until its concentration reaches a critical threshold, at which point division is triggered \cite{wood2015sizing,turner2012cell,rhind2021cell,keifenheim2017size}. Importantly, the production of these molecules must be coupled to size, otherwise only a timer mechanism is possible, as previously shown in models of bacterial size control \cite{elgamel2024effects,serbanescu2020nutrient,harris2016relative}. Indeed, for yeast, experiments show that the concentrations of various proteins scale with size \cite{chen2020differential,miller2023fission,bashir2023size}. Furthermore, recent experiments in fission yeast have shown that the concentrations of key proteins increase throughout the cell cycle in a size-dependent manner, rather than simply correlating with size \cite{bashir2023size}.

Surprisingly, experiments that altered size-dependent production of these proteins \cite{bashir2023size} or removed key proteins thought to act as size sensors in fission yeast \cite{wood2013pom1} revealed no impact on size control. Additionally, models relying on concentration accumulation to a threshold are unstable in principle, because division alone does not change concentrations, and therefore the next generation starts at the threshold immediately. This leaves the molecular mechanism responsible for yeast size control, and concentration-based control in general, widely unresolved.

Here, we introduce a general mechanistic model for concentration-based size control that relies on molecular concentration checkpoints for cell cycle progression. First, we demonstrate mathematically, without specifying a model, that to achieve sizer control through a concentration threshold, the concentration must be a pure function of size. Second, we show that a concentration-based mechanism for size control is only stable for multiple cell cycle checkpoints (stages). Third, we show that our model predicts the robustness of size control against disturbances in the production of molecules at an individual stage. Last, we compare this prediction, and the ensuing effects on the size statistics, to recent experimental data in fission yeast.

\begin{figure}
\includegraphics[width=\columnwidth]{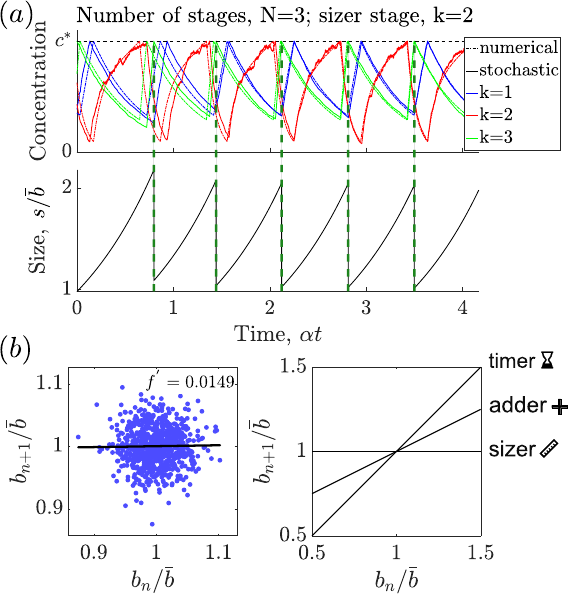}
\caption{(a) Between divisions, the cell cycle consists of multiple stages (three here). In each stage, one molecule must reach a critical concentration threshold to progress to the next stage while the remaining molecules are degraded. The final stage triggers division. The second stage was chosen as the sizer stage ($k=2$, red). Size grows exponentially. (b) (Left) The best fit line (black) for the simulation data of birth sizes shows a slope of $f^{'}=0.0149$, indicating sizer control ($f'=0$). Sizer control prevails as the dominant strategy when only one stage functions as a sizer. (Right) The slope of the $b_{n+1}$ vs $b_{n}$ map illustrates the size control strategy.}
\label{mdl}
\end{figure}

We start by deriving the functional form of the concentration required to achieve sizer control. In the $n$th generation, cell size is $s(b_n,t)$, where $b_n$ is birth size in that generation and $t$ is time since birth. In general, the concentration of a molecule can depend on $s$, $t$, and $b_n$. We therefore write the most general concentration function within a generation as $c(s,b_n,t)$. The explicit dependence on $b_n$ allows for transient molecular dynamics that depend on the initial size in that generation. Accounting for this transient dependence is important when molecular dynamics occur on a time scale comparable to size dynamics. Since $s$ is itself a function of $b_n$ and $t$, $c(s,b_n,t)$ is more generally written as $c(b_n,t)$. We note that our choice of the concentration to be a function of size and time does not exclude more complex dependencies on other molecular factors that can depend on size, time, or both. To trigger division, a concentration threshold, $c^*$, must be reached. At division, $t=T_n$, size and concentration satisfy the equations $s(b_n,T_n)=2b_{n+1}$ and $c(b_n,T_n)=c^*$, where the first equation assumes the cell divides in half. Taking the derivative of both equations with respect to $b_n$, we find
\begin{align}
	\frac{\partial s(b_n,T_n)}{\partial b_n} + \frac{\partial s(b_n,T_n)}{\partial T_n} \frac{\partial T_n}{\partial b_n}&=2f^{'},\label{vol_der}\\
    \frac{\partial c(b_n,T_n)}{\partial b_n} + \frac{\partial c(b_n,T_n)}{\partial T_n} \frac{\partial T_n}{\partial b_n}&=0,\label{conc_der}
\end{align}
where $f^{'}=\partial b_{n+1}/\partial b_n$. Stability requires $|f^{'}|<1$, with $f^{'}=0, 1/2$, and $1$ indicating sizer, adder, and timer control, respectively \cite{willis2017sizing}. A timer ($f^{'}=1$) has perfect birth size correlation across generations and is unstable due to its lack of robustness to noise, while a sizer ($f^{'}=0$) has no correlation across generations, a hallmark of sizer control. In our formalism, the concentration plays a similar role to a hazard rate, previously introduced for bacteria \cite{kennard2016individuality, grilli2017relevant}. In the case of a sizer, the hazard rate is independent of birth size, and division is triggered solely based on size \cite{grilli2017relevant}. Importantly, our formalism assigns the hazard rate an explicit molecular origin through the dynamics of the molecular species controlling division. For $-1<f^{'}<0$, size remains stable but fluctuations are overcorrected around the mean. Solving Eq.\ \eqref{vol_der} for $\partial T_n/\partial b_n$ and substituting in Eq.\ \eqref{conc_der} with $f^{'}=0$ yields
\begin{equation}
\label{ratios}
    \frac{\partial c(b_n,T_n)}{\partial b_n}\Big/ \frac{\partial c(b_n,T_n)}{\partial T_n}=\frac{\partial s(b_n,T_n)}{\partial b_n}\Big/ \frac{\partial s(b_n,T_n)}{\partial T_n}=\rho,
\end{equation}
where $\rho$ is specified by the size growth dynamics (e.g., linear or exponential). For linear growth, $s=\alpha T_n+b_n$, and  then $\rho=\frac{\partial s}{\partial b_n}\Big/ \frac{\partial s}{\partial T_n}=1/\alpha$, where $\alpha$ is the growth rate. Similarly, for exponential growth, $s=b_n e^{\alpha T_n}$, giving $\rho=1/\alpha b_n$. Assuming exponential growth, Eq.\ \eqref{ratios} becomes
\begin{equation}
\label{conc_sol}
    \frac{\partial c(b_n,T_n)}{\partial b_n}\Big/ \frac{\partial c(b_n,T_n)}{\partial T_n}=1/\alpha b_n,
\end{equation}
which can be solved using separation of variables \cite{supp} and yields $c(b_n,T_n)=a (b_n e^{\alpha T_n})^k=a s^k$, where $a$ and $k$ are constants. The general solution is the sum of all possible solutions, $c(b_n,T_n)=\sum_j^{\infty} a_j s^{k_j}$. This implies that any power series in size will satisfy Eq.\ \eqref{conc_sol}. Since any function of size can be expanded as a power series, we conclude that, to achieve sizer control, the concentration must follow a pure function of size, where pure means that all dependence on $b_n$ and $T_n$ must enter via $s$. This requires the concentration $c(b_n,t)$ to only depend on the birth size $b_n$ through how it depends on size, $c(b_n,t)=F(s(b_n,t))$. Intuitively, this means that thresholding $c$ is equivalent to thresholding $s$, since at the threshold $c^*$, size is $s^*=F^{-1}(c^*)$ on average. In deriving this general condition we have not specified whether size control is implemented through molecular concentration accumulation (as in fission yeast \cite{wood2015sizing,rhind2021cell}) or dilution (as in budding yeast \cite{rhind2021cell,schmoller2015dilution}) to the threshold. Thus, we expect this condition to apply in both cases. While simple, the condition $c(b_n,t)=F(s(b_n,t))$ is hard to implement dynamically, since cells do not directly control $c$, but $\dot{c}$. Therefore, for a growing cell with $\dot{c}\propto F(s(b_n,t))$, the relaxation time scale of $c$ must be much faster than the dynamics of $s$, such that $c$ quickly reaches an $s$-dependent quasi-steady state.

Up to this point, we have considered a single molecule that triggers division when its concentration reaches a fixed threshold. However, this division control mechanism is unstable, as the concentration is equal before and after division, which leads to multiple consecutive divisions. To address this issue, we introduce a model that relies on a multistage progression towards division. Serial models for cell-cycle progression were previously explored in the context of bacterial size control \cite{micali2018dissecting, adiciptaningrum2015stochasticity}. Each stage commences when the concentration of a specific molecule reaches a critical level [Fig.\ \ref{mdl}a]. Here, we focus on the case of molecular accumulation to a threshold, and we expect all results to hold in the case of molecular dilution. In our model, the cell cycle can consist of $N$ sequential stages, during each of which only one molecule is produced while all other molecules are degraded. Only the final stage triggers division. This approach allows the concentration of each molecule to fall below the threshold required in its respective stage, thereby preventing premature triggering of subsequent stages, including division. As a result, the stability of the lineage is maintained. Alternative implementations of a multistage cell cycle which lead to the same results include allowing all molecules to be produced throughout the cell cycle, while degradation takes place in the final stage. Throughout the paper we assume exponential size growth, and our results hold for linearly growing cells as well \cite{supp}.

The general model, for $N$ stages, is given by
\begin{equation}
\label{size}
\dot{s}=\alpha s,
\end{equation}
\begin{equation}
\begin{aligned}
\label{conc_dynamics}
\dot{c}_1&=\theta(0<t\leq T_1)(\mu_1 s+ \nu_1) - (\alpha + \lambda_1) c_1, \\	\dot{c}_2&=\theta(T_1<t\leq T_2)(\mu_2 s+ \nu_2) - (\alpha + \lambda_2) c_2, \\[-1ex] &\ \ \vdots \\ \dot{c}_N&=\theta(T_{N-1}<t\leq T_N)(\mu_N s+ \nu_N) - (\alpha + \lambda_N) c_N,
\end{aligned}
\end{equation}
where $c$ is molecular concentration, $\mu s$ accounts for size-dependent production, $\nu$ is a constant rate that accounts for size-independent production, $\alpha$ is the growth rate, $\lambda$ is the degradation rate, $\theta$ is the Heaviside step function defined as
\begin{equation}
\label{heaviside}
\theta(T_{j-1}<t\leq T_j)=
\begin{cases} 
      1 & T_{j-1}<t\leq T_j, \\
      0 & \text{otherwise}, 
\end{cases}
\end{equation}
and $T_j$ is the time at which the $j$th threshold is reached. Note that $\alpha c$ is an extra degradation term resulting from concentration dilution due to cell growth.

The assumption of multistage progression towards division is biologically well-supported, as the cell cycle transitions in eukaryotes are controlled by the concentrations of different molecular factors (such as Wee1 and Cdc25 in fission yeast \cite{keifenheim2017size}, Whi5 in budding yeast \cite{bertoli2013control}, and RB in mammalian cells \cite{giacinti2006rb,zatulovskiy2020cell}). Next, we demonstrate that the presence of a single molecule achieving sizer control in one stage is sufficient for the sizer strategy to dominate the entire cell cycle. This was previously demonstrated and termed the sizer theorem in \cite{micali2018dissecting}.

The slope of the discrete map of $b_{n+1}$ and $b_n$ indicates the size control strategy \cite{willis2017sizing}, illustrated in Fig.\ \ref{mdl}b. The model allows us to derive a general expression for the slope, given by
\begin{equation}
\begin{aligned}
\label{slope}
f^{'}=\frac{\partial b_{n+1}}{\partial b_n}=& \frac{1}{2}\Big(\frac{\partial s}{\partial b}- \frac{\partial s}{\partial t} \frac{\partial c_1}{\partial b}/\frac{\partial c_1}{\partial t}\Big)\Big|_{b=b_1, t=T_1}\\& \Big(\frac{\partial s}{\partial b}- \frac{\partial s}{\partial t} \frac{\partial c_2}{\partial b}/\frac{\partial c_2}{\partial t}\Big)\Big|_{b=b_2, t=T_2}\cdots \\& \cdots \Big(\frac{\partial s}{\partial b}- \frac{\partial s}{\partial t} \frac{\partial c_N}{\partial b}/\frac{\partial c_N}{\partial t}\Big)\Big|_{b=b_N, t=T_N}.
\end{aligned}
\end{equation}
Eq.\ \eqref{slope} is a general version of Eqs.\ \eqref{vol_der} and \eqref{conc_der} for a multistage model and is independent of size and concentration dynamics \cite{supp}. It only assumes the existence of stage-specific concentration thresholds. From Eq.\ \eqref{slope}, it becomes clear that if any molecule in any stage achieves sizer control, then, using Eq.\ \eqref{ratios}, $f^{'}=0$ and sizer control dominates the control strategy.

Within the model of Eq.\ \eqref{conc_dynamics}, sizer control is achieved in a given stage $k$ if the degradation rate of the produced molecule in that stage is much larger than the growth rate,  $\lambda_k \gg \alpha$. In this case, $c_k$ reaches quasi-steady state very quickly ($\dot{c}_k=0$), and from Eq.\ \eqref{conc_dynamics} we find $c _k \approx (\mu_k s+\nu_k)/(\alpha+\lambda_k)\approx (\mu_k s+\nu_k)/\lambda_k$. We see that the concentration is a pure function of size $s$, and therefore satisfies the general criterion for achieving sizer control in Eq.\ \eqref{ratios}. In Fig.\ \ref{mdl}, we simulated Eq.\ \eqref{conc_dynamics} for $N=3$, with the second stage, $k=2$, chosen as the sizer stage [Fig.\ \ref{mdl}a]; we find the slope of the map consistent with the sizer value $f^{'}=0$ [Fig.\ \ref{mdl}b]. This approach does not depend on the number of stages, the placement of the sizer stage within the cell cycle, or the specifics of the molecular dynamics within each stage, provided that the stage concentration threshold is reached and that at least one stage implements the sizer strategy.

\begin{figure}
\includegraphics[width=.93\columnwidth]{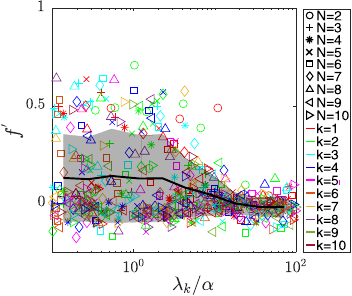}
\caption{Each point indicates the slope of the best fit line in the scatter plot of birth sizes, $b_{n+1}$ vs $b_n$. Each shape represents the number of stages before division, while the color represents the stage for which we track the degradation to growth rate ratio, $\lambda_k/\alpha$, on the $x$-axis. In all simulations, only $\mu_k$, $\nu_k$, and $\alpha$ were fixed. All other parameters were uniformly sampled in log space. When $\lambda_k/\alpha \gg 1$, at least one sizer stage exists and the simulation points collapse to $f^{'}=0$.}
\label{sizercollapse}
\end{figure}

We demonstrate the model's robustness for different numbers of stages, allowing different stages to assume the role of the sizer [Fig.\ \ref{sizercollapse}]. The parameters $\mu$, $\nu$, and $\lambda$ are uniformly sampled in log space. On the $x$-axis, we plot the ratio $\lambda_k/\alpha$ for stage $k$ while keeping $\mu_k$ and $\nu_k$ fixed. Our findings show a large cloud that spans many control strategies, depending on the sampled parameters. We expect that if only one stage serves as a sizer ($\lambda_k/\alpha\gg 1$), it dominates the control regardless of the strategy utilized in other stages. Indeed, we see the collapse of the simulation points to $f^{'}=0$ as $\lambda_k/\alpha$ increases. A similar plot is obtained for linear cell growth \cite{supp}. Biologically, this suggests that cells do not need to maintain stringent size control throughout the entire cell cycle to achieve sizer control overall. Rather, strong control over just one stage suffices, regardless of its order in the cell cycle. Fig.\ \ref{sizercollapse} also implies that size control is robust to perturbations to non-sizer stages. However, such perturbations could in principle affect cell size statistics, such as the mean and variance of size.

\begin{figure*}
\centering
\includegraphics[width=1\linewidth]{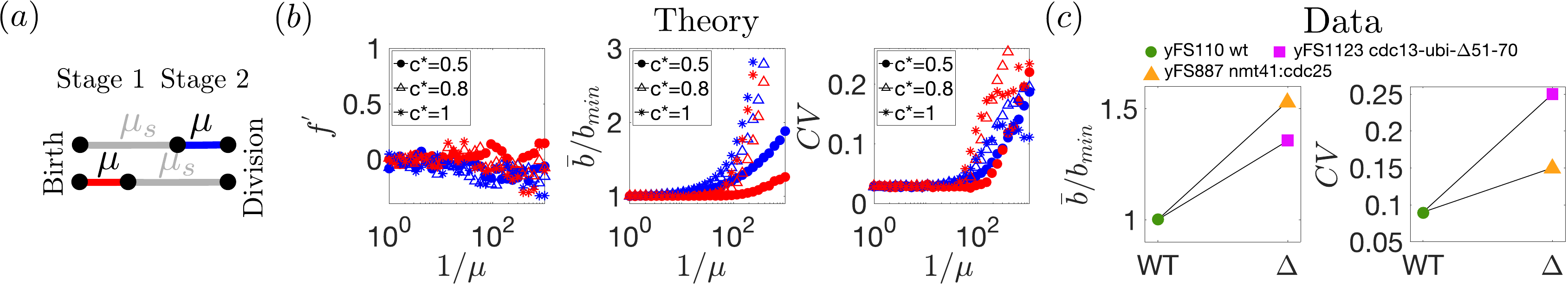}
\caption{Altering the size-dependent production of non-sizer stages does not affect size control. (a) A simple cell cycle model consists of only two stages. $\mu$ and $\mu_s$ indicate the size-dependent production of non-sizer and sizer molecules, respectively. The sizer stage takes up the majority of the cell cycle. The color indicates altered stages in the next panel. (b) Altering the size-dependent production of different stages does not halt size control, but affects size statistics. Sizer control is still achieved. Note the mean rescaled size ($\bar{b}/b_{min}$) and noise ($CV$) increase with decreased size dependence in molecular production (increased $1/\mu$) during the non-sizer stage. Results are shown for different concentration thresholds of the non-sizer molecules, $c^*$. The sizer molecule threshold is $c^*=10$, selected to ensure a longer sizer stage. (c) Experimental data, from ref. \cite{bashir2023size}, show that decreasing the size-dependent production of proposed size control proteins increases both the mean size and $CV$ at septation.}
\label{altersizer}
\end{figure*}

To investigate the effect of molecular perturbations on cell size statistics, we use a simple version of our model with only two stages, a sizer stage and a non-sizer stage [Fig.\ \ref{altersizer}a]. Then, we perturb the production of molecules by lowering their size-dependent production rate, $\mu$ (shifting towards timer). Finally, we plot the effects on size control, as well as mean size and noise (standard deviation over the mean, or coefficient of variation, $CV$). We find that size control is robust to perturbations as expected; $f^{'}$ values are at or near zero [Fig.\ \ref{altersizer}b]. However, mean size and $CV$ are affected significantly by perturbations, with both mean size and $CV$ increasing with lower size-dependent production. Intuitively, this is because lowering size-dependent production makes reaching the threshold more time consuming, resulting in larger sizes overall. Additionally, it turns molecules into timers, which is known to have strong size noise \cite{susman2018individuality,amir2014cell}. This indicates that control mechanisms in different stages can still have major implications either by producing very large cells or introducing strong size noise in the population.

Recent experiments in fission yeast have investigated how cell size is affected by perturbations of key cell cycle proteins Pom1 \cite{wood2015sizing}, Cdc13 \cite{bashir2023size}, and Cdc25 \cite{bashir2023size}. All are proposed to be responsible for size sensing and control, with Cdc13 and Cdc25 shown explicitly to have size-dependent production \cite{bashir2023size}. When the expression of these proteins was changed from size-dependent to size-independent \cite{bashir2023size}, or removed \cite{wood2015sizing}, size control was found to be unaffected, which was surprising and raised the question of whether they are truly responsible for size control. Our results provide a potential explanation: they may be involved in size control, but a different molecule takes the role of the sizer. Furthermore, when Cdc13 and Cdc25 were perturbed \cite{bashir2023size}, both the mean and CV of cell size at septation increased compared to the wild-type [Fig.\ \ref{altersizer}c], consistent with the predictions of our model [Fig.\ \ref{altersizer}b].


In this work, we established the general requirements for sizer control based on molecular concentration thresholds, which are utilized by eukaryotic cells, particularly yeast \cite{facchetti2017controlling,sauls2016adder,turner2012cell}. We have demonstrated that, to achieve sizer control, the concentration must be a pure function of size. This requirement is generic, irrespective of the mechanism. Moreover, we address the instability of concentration-based models by proposing that cells must follow a multistage progression toward division, in which a molecule in each stage reaches a concentration threshold. Interestingly, to achieve sizer control, only one stage is required to function as a sizer, previously termed the sizer theorem \cite{micali2018dissecting}. Our multistage model suggests that eukaryotes (with cell cycles stages) are more compatible with sizer-like control. This could explain why sizer control has only been reported in systems like fission yeast and is rarely observed in bacteria. It is important to point out that the most widely reported eukaryotic sizer is in fission yeast, \textit{S. pombe}. In other eukaryotes, such as budding yeast, distinct control strategies, particularly adder, have been reported \cite{chandler2017adder,soifer2016single}. Using simulations, we have shown that perturbations to non-sizer stages do not affect size control; however, size statistics are impacted. Our predictions for these impacts are consistent with recent experimental data. A limitation of our study is that we test the model predictions against only one experimental dataset. More experimental studies in known sizers are required to demonstrate the general applicability of the multistage model.

Our results suggest that cells have compensatory mechanisms for maintaining size control. It may be that size control is redundant and includes multiple size checkpoints throughout the cell cycle to protect size control against perturbations. Alternatively, the perturbed molecular factors may not be responsible for size control, despite being produced in a size-dependent manner. Therefore, size-dependent production is not sufficient alone to yield a sizer control mechanism. We predict that the sizer molecule acts through a separation of timescales in which the molecular degradation timescale is much faster than the size growth timescale. Experimentally, this can be tested by perturbing the degradation of known cell cycle control molecules, as has been done to investigate adder control in bacteria \cite{si2019mechanistic}. If size-dependent production and strong degradation are indeed the mechanisms by which cells sense and control size, confining size control to one cell cycle stage may be energetically more efficient, as strong degradation can be energetically costly \cite{lahtvee2014protein}. Our simple model coarse-grains many of the complexities of the size-control system, which is not entirely based on molecular concentrations \cite{facchetti2019reprogramming}. Nevertheless, we chose the simplest possible implementation of a sizer, motivated by experimental observations that molecular regulators that scale with size can contribute to size control \cite{keifenheim2017size,miller2023fission}.

The molecular mechanism behind size control in eukaryotes that exhibit sizer correlations, particularly fission yeast, remains an open question. We outline an experimental procedure based on our multistage model that can aid in the search for sizer molecules. Since our model assumes symmetric division, this procedure is only applicable to symmetrically dividing fission yeast, as division asymmetry affects division control in systems like budding yeast \cite{proulx2026division}. Our model predicts that birth size is forgotten abruptly at a particular cell cycle stage, not gradually throughout the cell cycle. This prediction can be tested by measuring the cross-correlation between the cell sizes at different time points during the cell cycle (i.e. measuring the correlation between $s(t_1)$ and $s(t_2)$, where $t_1$ and $t_2$ are different time points in the cell cycle). Correlation would be lost if the sizer molecule was activated between $t_1$ and $t_2$. This would confirm our prediction, identify the relevant cell cycle stage, and potentially identify the sizer molecule.

Understanding why different organisms employ one control mechanism over another is still a subject of ongoing research. While much has been explored regarding why systems like bacteria utilize the adder mechanism \cite{hobson2025evolutionary, si2019mechanistic, elgamel2024effects}, it remains underinvestigated why systems like fission yeast exhibit sizer control. Researching the effects of both mechanisms on population growth, function, evolution, and cell physiology is a compelling avenue of future inquiry \cite{rhind2021cell,genthon2025noisy,hobson2025evolutionary,jafarpour2019cell,jafarpour2023evolutionary,hein2024asymptotic}.


\begin{acknowledgments}
We thank Farshid Jafarpour for helpful suggestions. This work was supported by the National Science Foundation (Grant Nos.\ PHY-2118561 and DMS-2245816), the Emil Sanielevici Scholarship from the Department of Physics and Astronomy at the University of Pittsburgh, and the Andrew Mellon Predoctoral Fellowship from the University of Pittsburgh.
\end{acknowledgments}

\bibliography{refs}
\bibliographystyle{unsrt}

\onecolumngrid
\newpage
\beginsupplement

\begin{center}
{\bf SUPPLEMENTAL MATERIAL}
\end{center}
\section{Concentration dynamics}
Here, we solve Eq.\ \eqref{ratios} for both exponential and linear size growth. We focus first on the exponential size growth case ($s=b_ne^{\alpha T_n}, \rho=1/\alpha b_n$) where
\begin{equation}
\label{conc_sol_supp}
\frac{\partial c(b_n,T_n)}{\partial b_n}\Big/ \frac{\partial c(b_n,T_n)}{\partial T_n}=1/\alpha b_n.
\end{equation}
Assuming the solution takes the form $c(b_n,T_n)=A(b_n)B(T_n)$, where the $b_n$ and $T_n$ dependence can be separated. We find
\begin{equation}
\label{conc_sol_supp2}
\frac{b_n}{A}\frac{dA}{db_n}= \frac{1}{\alpha B}\frac{dB}{dT_n}=k,
\end{equation}
where $k$ is a constant. A and B are solutions of $\frac{b_n}{A}\frac{dA}{db_n}=k$ and $ \frac{1}{\alpha B}\frac{dB}{dT_n}=k$, respectively, solving them we find
\begin{align}
\label{sep_exp}
A&=c_1 b_n^k,\\
B&=c_2 (e^{\alpha T_n})^k,
\end{align}
therefore the solution is
\begin{equation}
\label{conc_supp4}
c(b_n,T_n)=a(b_n e^{\alpha T_n})^k=a s(b_n,T_n)^k,
\end{equation}
where $a$ is a constant and $c_1$ and $c_2$ were absorbed into $a$. We have identified $b_n e^{\alpha T_n}$ as size, evaluated at the division time $T_n$. Since $a$ and $k$ are arbitrary constants, the full solution is the sum of all possible values of $a$ and $k$. Thus, the full solution is given by
\begin{equation}
\label{conc_supp5}
c(b_n,T_n)=\sum_{j=0}^{\infty} a_j s^{k_j}.
\end{equation}
For linear size growth ($s=b_n+\alpha T_n, \rho=1/\alpha$), we have
\begin{equation}
\label{conc_sol_supp_lin}
\frac{\partial c(b_n,T_n)}{\partial b_n}\Big/ \frac{\partial c(b_n,T_n)}{\partial T_n}=1/\alpha,
\end{equation}
which leads to
\begin{equation}
\label{conc_sol_supp_lin2}
\frac{1}{A}\frac{dA}{db_n}= \frac{1}{\alpha B}\frac{dB}{dT_n}=k,
\end{equation}
which have the solutions
\begin{align}
\label{sep_lin}
A&=c_1 e^{k b_n},\\
B&=c_2 e^{k \alpha T_n}.
\end{align}
The solution is
\begin{equation}
\label{conc_supp_lin3}
c(b_n,T_n)=a e^{k (b_n + \alpha T_n)}=a e^{k s}.
\end{equation}
The full solution is the sum of all possible solutions,
\begin{equation}
\label{conc_supp_lin4}
c(b_n,T_n)=\sum_{j=0}^{\infty} a _j e^{k_j s}.
\end{equation}
We can expand the summand as a power series in $k_j s$ and find
\begin{equation}
\label{conc_supp_lin5}
c(b_n,T_n)=d_0+d_1 s + d_2 s^2 + d_3 s^3 +...=\sum_{i=0}^{\infty} d_i s^i,
\end{equation}
where $d_i = \sum_{j=0}^{\infty}\frac{1}{i!} a_j (k_j)^i$. In both the exponential and linear size growth cases, the final solution is a power series in size. This implies that the solution is any pure function of size, because it can be written as a power series in size. Therefore, the concentration has to be a pure function of size to achieve sizer control. 

\section{General $f^{'}$}
\begin{figure}
\includegraphics[width=0.5\columnwidth]{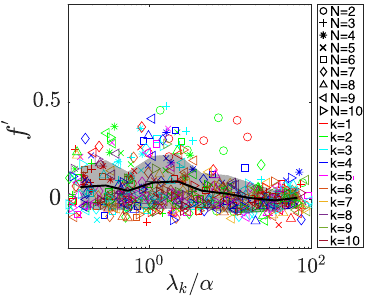}
\caption{Collapse plot with linear size growth. Each point indicates the slope of the best fit line in the scatter plot of birth sizes, $b_{n+1}$ vs $b_n$. Each shape represents the number of stages before division, while the color represents the stage for which we track the degradation to growth rate ratio, $\lambda_k/\alpha$, on the $x$-axis. In all simulations, only $\mu_k$, $\nu_k$, and $\alpha$ were fixed. All other parameters were uniformly sampled in log space. When $\lambda_k/\alpha \gg 1$, at least one sizer stage exists and the simulation points collapse to $f^{'}=0$.}
\label{linear_collapse}
\end{figure}

Assuming a general model with $N$ cell cycle stages, the concentrations of the $N$ cell cycle molecules at their thresholds are given by
\begin{equation}
\begin{aligned}
\label{conc_th}
c_1^*&=c_1(b_1,T_1), \\	c_2^*&=c_2(b_2,T_2), \\[-1ex] &\ \ \vdots \\ c_N^*&=c_N(b_N,T_N),
\end{aligned}
\end{equation}
where $c_N^*, b_N$, and $T_N$ are the concentration threshold, initial size, and end time of the $N^{th}$ stage, respectively. Differentiating Eq.\ \eqref{conc_th} with respect to the birth size $b_n$ gives
\begin{equation}
\begin{aligned}
\label{conc_th_2}
		\frac{\partial c_1}{\partial b_1} &+ \frac{\partial c_1}{\partial T_1} \frac{\partial T_1}{\partial b_n}=0,  \\
		\frac{\partial c_2}{\partial b_2} \frac{\partial b_2}{\partial b_n} &+ \frac{\partial c_2}{\partial T_2} \frac{\partial T_2}{\partial b_n}=0,  \\[-1ex] &\ \ \vdots \\
		\frac{\partial c_N}{\partial b_N} \frac{\partial b_N}{\partial b_n} &+ \frac{\partial c_N}{\partial T_N} \frac{\partial T_N}{\partial b_n}=0,
\end{aligned}
\end{equation}
where the derivatives of the concentration thresholds are zero because they are constant, and $b_1=b_n$. The initial sizes are defined by
\begin{equation}
\begin{aligned}
\label{size_th}
		b_2&=s(b_n,T_1), \\	b_3&=s(b_2,T_2), \\[-1ex] &\ \ \vdots \\ 2 b_{n+1}&=s(b_{N},T_{N}),
\end{aligned}
\end{equation}
where we used the fact that $b_1=b_n$ and $b_{N+1}=2 b_{n+1}$. Eq.\ \eqref{size_th} can be differentiated with respect to $b_ n$ and gives
\begin{equation}
\begin{aligned}
\label{size_th_2}
		\frac{\partial s}{\partial b_n} &+ \frac{\partial s}{\partial T_1} \frac{\partial T_1}{\partial b_n}=\frac{\partial b_2}{\partial b_n},  \\
		\frac{\partial s}{\partial b_2} \frac{\partial b_2}{\partial b_n} &+ \frac{\partial s}{\partial T_2} \frac{\partial T_2}{\partial b_n}=\frac{\partial b_3}{\partial b_n},  \\[-1ex] &\ \ \vdots \\
		\frac{\partial s}{\partial b_{N}} \frac{\partial b_{N}}{\partial b_n} &+ \frac{\partial s}{\partial T_{N}} \frac{\partial T_{N}}{\partial b_n}=2 f^{'},
\end{aligned}
\end{equation}
where $f^{'}=\frac{\partial b_{n+1}}{\partial b_n}$. Solving for $\frac{\partial T_1}{\partial b_n}, \frac{\partial T_2}{\partial b_n}, \hdots,$ and $\frac{\partial T_N}{\partial b_n}$ using Eqs.\ \eqref{conc_th_2} and substituting into Eqs.\ \eqref{size_th_2} yields
\begin{equation}
\begin{aligned}
\label{size_th_3}
		\frac{\partial s}{\partial b_n} &- \frac{\partial s}{\partial T_1} (\frac{\partial c_1}{\partial b_1} / \frac{\partial c_1}{\partial T_1})=\frac{\partial b_2}{\partial b_n},  \\
		\frac{\partial s}{\partial b_2} \frac{\partial b_2}{\partial b_n} &- \frac{\partial s}{\partial T_2} (\frac{\partial c_2}{\partial b_2} \frac{\partial b_2}{\partial b_n}/ \frac{\partial c_2}{\partial T_2})=\frac{\partial b_3}{\partial b_n},  \\[-1ex] &\ \ \vdots \\
		\frac{\partial s}{\partial b_{N}} \frac{\partial b_{N}}{\partial b_n} &- \frac{\partial s}{\partial T_{N}} (\frac{\partial c_N}{\partial b_N} \frac{\partial b_N}{\partial b_n}/ \frac{\partial c_N}{\partial T_N})=2 f^{'}.
\end{aligned}
\end{equation}
We substitute $\frac{\partial b_2}{\partial b_n}$ into the second equation, then $\frac{\partial b_3}{\partial b_n}$ into the third equation, and so on until the final equation in the chain. Eventually, we obtain
\begin{equation}
\begin{aligned}
\label{slope_full}
f^{'}= \frac{1}{2}\Big(\frac{\partial s}{\partial b_n}- \frac{\partial s}{\partial T_1} \frac{\partial c_1}{\partial b_1}/\frac{\partial c_1}{\partial T_1}\Big) \Big(\frac{\partial s}{\partial b_2}- \frac{\partial s}{\partial T_2} \frac{\partial c_2}{\partial b_2}/\frac{\partial c_2}{\partial T_2}\Big)\cdots \Big(\frac{\partial s}{\partial b_N}- \frac{\partial s}{\partial T_N} \frac{\partial c_N}{\partial b_N}/\frac{\partial c_N}{\partial T_N}\Big).
\end{aligned}
\end{equation}
Throughout the derivation we did not specify the size growth dynamics. Thus, regardless of the growth dynamics, we expect sizer control to dominate the control strategy if one stage achieves sizer control. Indeed, we obtain a collapse plot of the control strategy for linear growth, similar to the one shown for exponential growth in the main text [Fig.\ \ref{linear_collapse}]. Note the decreased range of $f^{'}$ in Fig.\ \ref{linear_collapse}. This is due to the fact that timer control is equivalent to adder control for linearly growing cells. 

\section{Stochastic simulations}
To simulate the concentration dynamics we used the stochastic simulation (Gillespie) algorithm. We first need to derive the equivalent molecule number, $x$, dynamics using
\begin{equation}
\label{ctox}
\dot{x}=\frac{d(cs)}{dt}=s\frac{dc}{dt}+c\frac{ds}{dt}=s\dot{c}+c\dot{s}.
\end{equation}
We get
\begin{equation}
\begin{aligned}
\label{num_dynamics_supp}
\dot{x}_1&=\mu_1 s^2+ \nu_1 s - \lambda_1 x_1, \\	\dot{x}_2&=\mu_2 s^2+ \nu_2 s - \lambda_2 x_2, \\[-1ex] &\ \ \vdots \\ \dot{x}_N&=\mu_N s^2+ \nu_N s - \lambda_N x_N.
\end{aligned}
\end{equation}
Then, we use Eqs.\ \eqref{num_dynamics_supp} to simulate molecule number dynamics. The molecule number dynamics are independent of size growth dynamics (linear or exponential). The transition rates of the reactions are
\begin{equation}
\begin{aligned}
\label{trans_prob}
T^+_{x_1\rightarrow x_1+1}&=\mu_1 s^2, T^+_{x_1\rightarrow x_1+1}=\nu_1 s, T^-_{x_1\rightarrow x_1-1}=\lambda_1 x_1, \\	T^+_{x_2\rightarrow x_2+1}&=\mu_2 s^2, T^+_{x_2\rightarrow x_2+1}=\nu_2 s, T^-_{x_2\rightarrow x_2-1}=\lambda_2 x_2, \\[-1ex] &\ \ \vdots \\ T^+_{x_N\rightarrow x_N+1}&=\mu_N s^2, T^+_{x_N\rightarrow x_N+1}=\nu_N s, T^-_{x_N\rightarrow x_N-1}=\lambda_N x_N.
\end{aligned}
\end{equation}
Then concentrations are easily obtained by dividing by volume after each reaction. The parameters change in each stage to specify which molecule is produced while the others are degraded. The sizer stage is determined by strong degradation.

\end{document}